\documentclass[12pt]{article}
\usepackage{epsfig,amssymb}
\newcommand{\nn}{\nonumber}

\newcommand{\be}{\begin{equation}}
\newcommand{\ee}{\end{equation}}
\newcommand{\bea}{\begin{eqnarray}}
\newcommand{\eea}{\end{eqnarray}}

\begin{document}
\setlength{\parskip}{0.45cm}
\setlength{\baselineskip}{0.75cm}
\begin{titlepage}
\begin{flushright}
CERN-TH/2000-024\\
FTUAM-00-02\\
IFT-UAM/CSIC-00-03\\
{\tt hep-ph/0001335} \\
\end{flushright}
%
\begin{center}
\Large
{\bf Fat brane phenomena}\\
\vspace{1.5cm}
\large
\centerline{ A.~De~R\'{u}jula$^{\rm a}$\footnote{derujula@nxth21.cern.ch}, 
A.~Donini$^{\rm b}$\footnote{donini@daniel.ft.uam.es}, 
M.B.~Gavela${\rm ^b}$\footnote{gavela@mail.cern.ch} 
and S.~Rigolin$^{\rm b}$\footnote{rigolin@mail.cern.ch}}
\begin{center}
\small{
(a) {\it Theoretical Physics Division, CERN, CH-1211 Geneva 23, Switzerland} \\
(b) {\it Departamento de F\'{\i}sica Te\'orica C-XI, Universidad Aut\'onoma de 
         Madrid, Cantoblanco, 28049 Madrid, Spain.}
}
\end{center}
\vspace*{0.5cm}
\normalsize
\large
{\bf Abstract} \\
\end{center}
\vspace*{-0.3cm}

\noindent{Gravitons could permeate extra space dimensions
inaccessible to all other particles, which would be confined to ``branes''.
We point out that these branes could be ``fat'' and have a non-vanishing
width in the dimensions reserved for gravitons. In this case the other
particles, confined within a finite width,
should have ``branon'' excitations. Chiral fermions behave differently
from bosons under dimensional reduction, and they may 
--or may not-- be
more localized than bosons. All these possibilities are
in principle testable and distinguishable, they could yield spectacular
signatures at colliders, such as the production of the first branon
excitation of $\gamma$'s or $Z$'s, decaying into their ground state
plus a quasi-continuum of graviton recurrences.
We explore these ideas in the realm of a future lepton collider
and we individuate a {\it dimensiometer}: an observable that would cleanly diagnose the number
of large ``extra'' dimensions.}

\today

\end{titlepage}
\newpage
\normalsize

No evidence counters the observation that we live in 3+1 dimensions.
Yet, a large fraction of the current theoretical-physics literature  
deals with extra space dimensions. 
Clearly, new dimensions must be different from
the ``old'' ones, the simplest possibility --of which the earliest
milestone~\cite{KK}
is due to Kaluza and Klein-- being to compactify them to a domain
of minute size. The theoretical interest in extra-dimensional
physics is kindled by successive ``superstring revolutions'',
which have ingrained the belief that there should be a total of
9 space dimensions.

An interesting remark~\cite{RSh,HW} is that different particles may move in 
different spaces; in particular gravity could permeate dimensions
into which quanta of other fields cannot propagate.
It is not excluded that these latter dimensions be large enough for
deviations from Newton's law to be observable at submillimetre distances.
It is also not excluded that the rest of the assumed 9 dimensions
be compactified in manifolds of $\cal{O}$(1) TeV$^{-1}$ size, which could 
make their effects testable at future colliders. These two 
remarks~\cite{Antoniadis:1990ew,Savas,Lykken}, however contrived,
have led to a surge of phenomenological interest~\cite{Review} in
new dimensions much larger than the tiny, gravitationally ``natural'', 
Planck length. We shall refer to the submillimetre and inverse-TeV 
dimensions as {\it large} and {\it small}, respectively.

Theoretical descriptions of the possible phenomenological consequences
of extra dimensions mix old concepts of compactification, such as towers
of Kaluza--Klein (KK) excited particles, with novel ones, such as
twisted sectors, D-branes, orientifolds, etc. We shall illustrate
how some of the key assumptions in these constructions translate into
observational tests, mainly in the form of selection rules.

In Type I string theories, ``Dirichlet $p$-branes'' or {\it D-branes},
are defined as $p$-dimensional spaces ($p\leq 9$) to which the ends of 
open strings attach \cite{Polchinski}. As an example, ordinary 3-space 
could be a 3-brane, to which all particles but (closed-string)
gravitons would be confined by the aforementioned boundary conditions. 
The space spanned by the dimensions where only gravitons propagate is 
called {\it the bulk}, its dimensionality is $\delta=6$ in the above 
example. One can also choose $\delta<6$ extra large dimensions,
by adopting a $p$-brane with $p=3$ ordinary  plus $6-\delta$ small
dimensions.

Branes should be the vacua of some so far unresolved string dynamics 
\cite{Ibanez}; they are hypersurfaces with a finite tension, $f$, and 
perhaps --like solitons-- with a finite extension. The question of the 
extension or ``width'' of a brane (into the directions orthogonal to it)
is obscured by the dualities of string dynamics: as for a monopole, what
may look like a composite object in one realization of the theory may be more
singular or ``elementary'' in a different one. Our intuition is that  
any of the objects that are solutions of a theory as non-singular as string theory 
ought to be non-singular: branes should have --in an operational sense to be 
defined anon-- a finite width $L$ of the order of $1/f$. The notion that 
branes are wide may be right or wrong, but it is testable in principle.

In attempts to bridge the gap between particle physics and string theory,
complex brane patterns have been considered: brane intersections, 
layered structures of parallel branes, etc. \cite{Dienes,Arkani}. We shall 
abstract from these {\it designer branes,} again, only the hint that these 
constructs may represent objects with a non-vanishing width~\cite{Bando}.

Consider first for illustration the case of one extra large dimension, $y$,  
compactified on a circle of radius $R$. The  coordinate $y$ represents 
the position of the ordinary-space 3-brane in the compact dimension. 
The arbitrariness of this position reflects a spontaneous symmetry 
breakdown, and implies that $y$ is a dynamical Goldstone field~\cite{sundrum1,us}.
Develop to first order the (1+4)-dimensional metric, assumed to be 
approximately flat, as $G=\eta+H/M^{3/2}$ (with $M^{3/2}=M_P/\sqrt{2\pi R}$ 
in terms of the conventional Planck mass).
The  tensor $H_{MN}$ consists of three parts: a 4-d graviton
$H_{\mu\nu}$, a graviphoton $H_{\mu 5}$ and a graviscalar $H_{55}$,
only the first one of which~\cite{add2,Giudice,Peskin1} 
will concern us here\footnote{The
graviphoton spouses the Goldstone field $y$ to acquire a 
mass~\cite{sundrum1,add2}; its coupling to ordinary matter involves the 
emission of a {\it phonon} (a local brane excitation) and is suppressed by 
two inverse powers of the brane's tension~\cite{us}.
The graviscalar is given a mass by whatever mechanism stabilizes the
compactified radius and  its couplings to ordinary matter are  
suppressed by mass factors~\cite{Giudice} and by the four inverse  
powers of the brane tension associated with double-phonon emission~\cite{us}.
Signatures of graviphotons and graviscalars are accordingly damped.}. 
In the familiar way, $H(x,y)$ can be expanded as a KK tower
of 4-d fields. For the graviton and its excitations,
\begin{equation}
H_{\mu\nu}(x, y) = \frac{1}{\sqrt{2 \pi R}}
\sum_{k=-\infty}^{k=\infty}  e^{i k y/R }\; H_{\mu\nu}^k (x)\, .
\label{KKGrav}
\end{equation}
The fifth-derivative in the kinetic terms of the Lagrangian results in a mass
$m_k=|k|/R$ for the $k$-th graviton recurrences. 
The mass gap is in the meV range for $R$ in the submillimetre range.
Conservation of momentum in the $y$ direction implies a selection
rule: a trilinear coupling of the $l,m,n$ gravitons vanishes unless
$l+m+n=0$, as the wave functions of Eq.~(\ref{KKGrav}) dictate.

In studying the couplings of the KK-excited gravitons to the standard
particles it is customary to place the latter on a {\it thin brane} of width $L=0$.
In the 5-dimensional illustration of the previous paragraph 
this corresponds to
setting $\Phi(x,y)\!\to\!\Phi(x)\delta(y)$ for all standard fields.
The specific positioning of the brane in the extra dimension is a 
breakdown of $y$-translational symmetry, or 5th momentum  
conservation, as a consequence of which the trilinear vertices 
$\Phi \Phi H^k$ for the emission of a graviton do not vanish,
even if $k$ (the overall ``lost'' 5th-momentum ) is not zero. 
This has extremely interesting phenomenological 
consequences~\cite{Giudice,Peskin1}. For a {\it fat brane}
whose profile is not a delta function, other selection rules are also
broken, allowing for the existence of couplings with even more
interesting outcomes.

Let ordinary space be a  fat 3-brane of width $L$ in the $y$ direction.
To construct a field-theoretical toy model~\cite{Arkani} of fields attached 
to this brane, consider a massless scalar field $\phi(x,y)$ confined  
to the interval $y\!=\!0$ to $y\!=\!L$, with periodic
boundary conditions. This field can be expanded as
$\phi(x,y)\!=\! \sum \phi^n (x) \varphi_n (y)$, with $\varphi_n (y)$
the stationary eigenfunctions that include a 
massless ground state, which are even in $y$:
\bea
\varphi_n (y) &=& \sqrt{ \frac{2 - \delta_{n0}}{L} } \,
\cos \left [ \frac{2 n \pi}{L} y  
\right ],
\qquad n \geq 0.
\eea
The (1+3)-dimensional quanta of
$\phi^n (x)$ have masses $m_n = 2\, n\, \pi /L$.
A confining brane acts as a compact dimension in that it results in these
KK-like states~\cite{Arkani}, which disappear in the $L\to 0$ limit.
We see no way, in a field-theoretical realm, to have confinement
to a finite-size domain without generating 
such states, to which we shall refer as {\it branons}, to distinguish them from
genuine KK excitations, whose wave function extends over the
whole of the compact dimension, as in Eq.~(\ref{KKGrav}).

\begin{figure}[t]
\begin{tabular}{cc}
\hskip -0.5truecm 
\epsfig{file=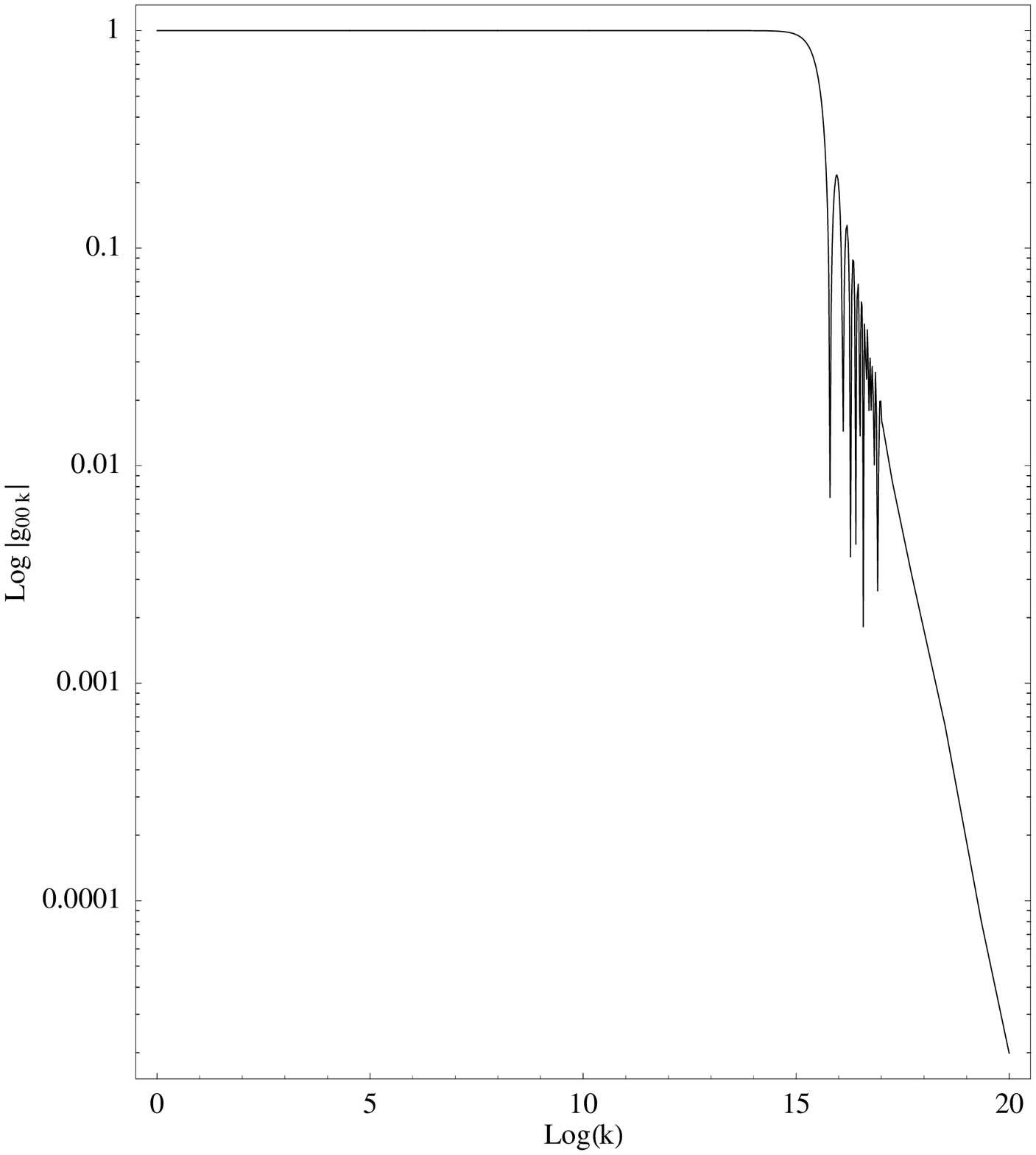, width=7.1cm} &
\hskip 0.5truecm
\epsfig{file=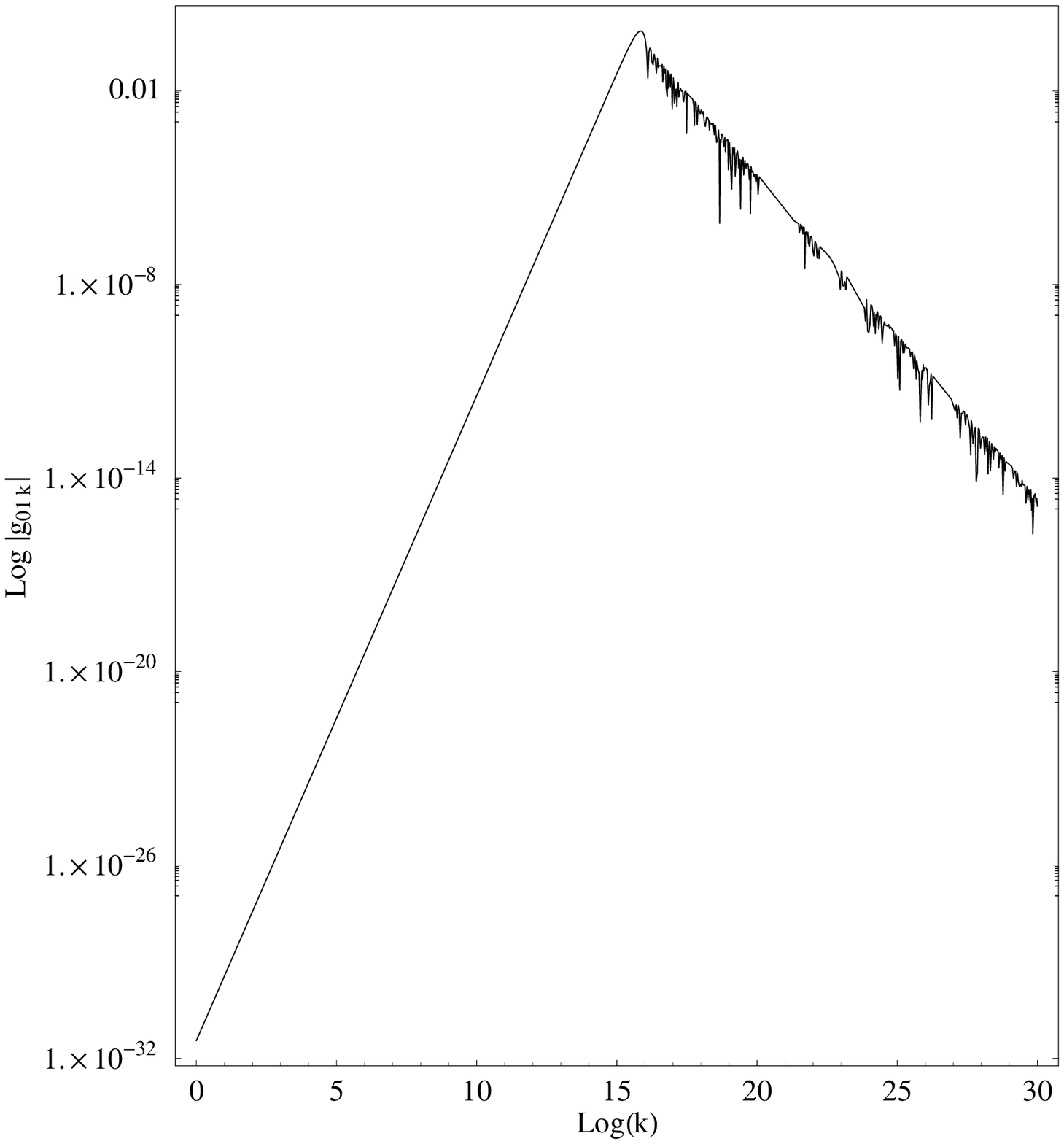, width=7.4cm} \\
\hskip 0.2truecm
{\small (a)}            &
\hskip 1.4truecm
{\small (b)}
\end{tabular}
\caption{{The couplings $ \log |g_{00k} |$ and $ \log |g_{01k} |$   
(with $g$ in units of $M_P^{-1}$) as functions of $\log k$, 
for $R=0.2$ mm and $L=1$ TeV$^{-1}$. The sharp dips in the
curves are bad renderings of a function with narrowly spaced zeros.}}
\label{fig:pllog0ik}
\end{figure}

The interaction Lagrangian of the scalar fields $\phi^n$ with the
gravitons $H^{(k)}$ is:
\be
\label{intHphi}
{\cal L}_{H \phi \phi} (x)  =
- \frac{1}{2} ~\sum_{kmn}\, g_{mnk}\, H^{(k)}_{\mu\nu} (x)\,
                 \,   \partial^\mu \phi^m (x) \,\partial^\nu \phi^n (x)\, ,
\ee
where the coupling $g_{mnk}$ is:
\be
g_{mnk} = \frac{1}{ M^{3/2}\, L\,\sqrt{2\pi R} } \int_0^{L} dy \;
\varphi_m(y)\, \varphi_n(y)\;e^{i k y/R }~.
\label{gkmn}
\ee
The amplitude for the emission of
the $k$-th KK mode of the
graviton by the zero mode of the scalar field is:
\be
g_{00k} = \frac{i}{ M_P } \,
\frac{ \xi \left ( 1 - e^{ i \frac{k }{ \xi } } \right ) }{ k}\, ,
\label{diagonal}
\ee
where $\xi\equiv R/L$. As an example, let $R=0.2$ mm and $L=1$ TeV$^{-1}$,
so that $\xi\simeq 10^{15}$. In Fig.~\ref{fig:pllog0ik}(a) we show the 
``elastic form factor'' $g_{00k}$
as a function of $k$. For the many modes for which $k\ll\xi$,
or in the thin-brane limit, the
coupling strength coincides with the universal
amplitude $\sqrt{G_N}=M_P^{-1}$ of the massless
graviton. For $k={\cal O}(\xi)$ or greater, the gravitons can discern  
the brane-size
structure of the scalar mode, and the form factor decreases.

Non-diagonal couplings are particularly interesting. Consider $g_{01k}$,
the gravitons' coupling to the fundamental and first-excited scalar branons:
\be
g_{01k} = \frac{ i \sqrt{2} }{ M_P } \,
\frac{ \xi k \left ( 1 - e^{ i \frac{k}{ \xi } } \right ) }{ k^2  
 - ( 2 \pi \xi)^2 }~.
\label{nondiagonal}
\ee
In  Fig.~\ref{fig:pllog0ik}(b) we show the ``transition form factor'' $g_{01k}$
as a function of $k$.
For $k\ll\xi$, $g_{01k}$ vanishes as $k^2$, the
graviton has too little 5-th momentum to undo the orthogonality of
$\phi^0$ and $\phi^1$. For $k={\cal O}(\xi)$ this selection rule is avoided, as the 
$k$-th graviton can absorb the momentum
inbalance. For even larger $k$ the form factor effect takes over and
$g_{01k}\!\to\! 0$. The explicit expressions in
Eqs.~(\ref{diagonal}) and (\ref{nondiagonal}) are specific to our toy model
of confinement to a brane, but their general behaviour
is the one to be intuitively expected: it
would be quite similar in any other simple model. 
The fact that $g_{01k}\!\propto\! k^2$ and $g_{00k}\!\propto\! k^0$
at low $k$ is general: it can be derived in an ``effective field-theory''
sense. These non-trivial
form factors have interesting  consequences,
as we shall see.

The fermions of the Standard Model occur in chiral multiplets.
It is notoriously difficult to generate these objects in field-theoretic
models of dimensional compactification. In heterotic string theories 
it is customary to face this problem by locating fermions in
the singular points of orbifolds (such as $y\!=\!0,\pi$
in a circle $0\!<\!y\!\leq\! 2\pi$ subject to the identification
$y\leftrightarrow -y$). In generalizations to brane constructs,
the role of the orbifold singularities is played by intersections
of branes. We abstract from all this the possibility
that  fermions be confined to spaces more singular
than the corresponding ones for bosons, as is often
done in the literature without further ado~\cite{Anton}. We shall,
in turn,
study this possibility and its negation. Reconsider the scalar field
attached to a fat brane that we have
discussed,  $\phi(x,y)$, and its trilinear coupling $\bar \psi   
\psi \phi$ to a
fermion  that is placed at a specific $y$-location: $\psi(x,y) \propto  
\psi(x)\delta(y)$. As in the case of graviton's KK--recurrences,
this breakdown of $y$-translational symmetry 
implies that there are no selection rules in the coupling $\bar \psi \psi  
\phi^n$: it does not vanish for branons with $n\neq 0$. 
This opens up an avenue to a very
rich phenomenology, as we proceed to discuss in the specific case of
QED on a fat brane.

The minimum number of extra large dimensions of conceivable empirical  
interest is\footnote{The radius of a single extra ``large'' dimension
would be unacceptably large~\cite{Savas}.} $\delta=2$.
This number is also the most amenable to experimental  
escrutiny and, in most of what follows, we concentrate on it.  
Paraphrasing~\cite{sundrum1}, 
we introduce a 6-vector $Y^M$, the first 4 of whose entries are the ordinary  
coordinates $x$ (with indices $\mu$, $\nu$,...) while the remaining two,  
labelled $y$ (with indices $a$, $b$,... for $M=5,6$) are the Goldstone fields  
specifying the position of the 3-brane in the bulk. On the brane, the 6-d
metric $G^{MN}$ induces a 4-d metric:
\be
g_{\mu \nu} (x,y) = G^{MN} (x,y) ~\partial_\mu Y^M (x)~ \partial_\nu Y^N (x)\; .
\ee
The brane action is akin to the one studied in~\cite{sundrum1} 
but for the fact 
that we use a fat brane with a shape $B(y)$:
\be
S = \int d^4 x \, d^2 y ~\sqrt{ - g}~
B(y)~[ - f^4 +  g_{\mu \nu} (x,y) \;
T^{\mu\nu} (x,y) ] \,+ \, ... 
\label{sqed}
\ee
where the brane's tension $f$ induces an energy density profile
$f^4\,B(y)$. The ellipsis in Eq.~(\ref{sqed}) stands for
higher-order terms in  $(\partial Y/f)$ and/or $L/R$, resulting from the extra-dimensional 
components of the energy--stress tensor. They are associated to phonon emission, 
and suppressed in comparison with the effects we study.  

We begin by discussing the $\delta\!=\!2$ fat-brane scenario with bosons 
confined to a domain of width $L$. Let $B(y)$, once again, 
be simply modelled as a square box. Fermions, as we discussed, 
will first be treated more singularly: $B(y)\to\delta(y)$. 
The fat-brane QED action (neglecting all terms involving phonons) is then:
\be
S_{QED}\! =\! \int \! d^4 x \, d^2 y \sqrt{ - g}\,B(y)\,g^{\mu \nu} \Big\{
i \bar \psi (x) \gamma_\mu D_\nu \psi (x)\,\delta(y)
+ \frac{1}{4}\; F_{\mu M} (x,y) {F^M}_\nu (x,y) \Big\}
\label{esta}
\ee
where $D_\mu = \partial_\mu - i e_6 A_\mu (x,y)$;
and $e_6=e\,L$: the 6-d electromagnetic coupling.

We KK-expand $G^{MN}$ as in
Eq.~(\ref{KKGrav}). We also develop the photon and photoscalars into   
towers of branons:
\be
A_M (x,y) = A_M^{(\vec{0})} (x) + \sqrt{\frac{2}{L}}\;
\sum_{\vec{n}} \cos\left[{ \frac{2 \pi}{L}\, \vec{n} \cdot \vec{y} }\right]
A_M^{(\vec{n})} (x)
\;\;\;\;\;\; n_i\geq 0 , 
\label{photons}
\ee
where $\vec{n} = (n_1, n_2)$. 
Half of the 4-d scalar components of the photon $A_M$ are gauge artefacts, 
which the gauge condition $\partial_M A^M (x, y) = 0$ eliminates. 
The 4-d excited modes of the photon (the branons) acquire 
mass through a Higgs mechanism that removes
from the spectrum all photoscalars other than the zero-modes. 
All in all, the 6-d fields $A_M$ result in a 4-d massless photon, 
its massive vector excitations 
and two massless photoscalars (one for each extra dimension). 

The branon mass splitting is $m = 2 \pi /L$. The lower bounds on 
the masses of excited photons
and Z's are of ${\cal O}$(1 TeV). The next generation of 
colliders can only hope 
to see or intuit the lowest excitations, to which we restrict 
the remaining discussion. Rotate the massive photons,
$A_\mu^{(\pm)}\!=\!(A_\mu^{(1,0)} \pm A_\mu^{(0,1)})/\sqrt{2}$,
and rescale the zero-mode photoscalars $A_a^{(\vec{0})} \to \sqrt{2}  
A_a^{(\vec{0})}$ in Eq.~(\ref{esta}), to obtain:
\bea
S_{QED} & \simeq & \int d^4 x 
\Big \{
i \bar \psi \gamma^\mu \partial_\mu \psi +
e \bar \psi \gamma^\mu
\left [ A_\mu^{(\vec{0})} + \sqrt{2} A_\mu^{(+)} \right ] \psi  \nn \\
&-& \sum_{i = \vec{0},+,-} \frac{1}{4}\, F_{\mu\nu}^{(i)} F^{\mu\nu (i)}
+ \frac{1}{2} \,\partial_\mu A_a^{(\vec{0})} \partial^\mu A_a^{(\vec{0})} \nn \\
&-& \frac{1}{4}\, m^2 \sum_{i = +,-} A_\mu^{(i)} A^{\mu (i)} 
- \frac{1}{M_P} \sum_{\vec{k}} H^{\mu\nu (\vec{k})}
T_{\mu\nu}^{(\vec{k})} \Big \} \, . 
\label{SQED}
\eea
The gravitational couplings of the zero and first-excited  modes
can be read from
\bea
T_{\mu\nu}^{(\vec{k})} &=&
i \bar \psi \gamma_\mu \partial_\nu \psi +
e \bar \psi \gamma_\mu \left ( A_\nu^{(\vec{0})} + \sqrt{2}  
A_\nu^{(+)} \right ) \psi \nn \\
&+& \frac{1}{4} \,\,
g_0^{(k_1, k_2)} \left (
F_{\mu \rho}^{(\vec{0})} {{F^\rho}_\nu}^{(\vec{0})} +
2 \partial_\mu A_a^{(\vec{0})} \partial_\nu A_a^{(\vec{0})} \right ) \nn \\
&+& \frac{1}{2 \sqrt{2} } \,\,
(g_1^{(k_1, k_2)} \pm g_1^{(k_2, k_1)} )
F_{\mu \rho}^{(\vec{0})} {{F^\rho}_\nu}^{(\pm)} \nn \\
&-& i \sqrt{2}\; m  
\left ( g_1^{(k_1 ,k_2)} \partial_\mu A_1^{(\vec{0})} \pm
        g_1^{(k_2, k_1)} \partial_\mu A_2^{(\vec{0})} \right )  
A_\nu^{(\pm)} \, ,
\label{Tmunu}
\eea
where we have defined $g_i^{(k_1,k_2)}\!=\!M_P\, g_{0 i k_1}\,g_{0 0 k_2}$,
to extend the $\delta\!=\!1$ results of Eqs.~(\ref{diagonal}) and (\ref{nondiagonal}) 
to one more extra dimension.

Following the weird but deeply rooted habit of denoting many fields
differently from their corresponding quanta, let $G^{(\vec k)}$ be the
gravitons of the field $ H^{(\vec k)}$, $\gamma$ and $\gamma^*$
be the photons of the fields $A_\mu^{(\vec{0})}$ and $A_\mu^{(+)}$,
and $\tilde\gamma_a$, with $a\,=\,1,2$, the photoscalars of the
fields $A_a^{(\vec 0)}$.
Fermions, as in Eq.~(\ref{SQED}), couple to $\gamma$ and $\gamma^*$,
but not to the odd massive vector mode, which will play no role here,
 nor to the massless photoscalars $\tilde\gamma$, for which a minor
role is reserved\footnote{Even if the scalar fields are massless,
the generalization of Eq.~(\ref{Tmunu}) to QCD does not entail a
 violation of the equivalence principle, since there is
no single-scalar emission by quarks or gluons.}.
The $\gamma\gamma G^{(\vec k)}$ vertex in Eq.~(\ref{Tmunu}),
relative to its standard gravitational strength in the  thin-brane 
limit~\cite{Giudice,Peskin1}, is reduced for a fat brane by a form factor 
$ g_0^{(k_1,k_2)}$. More interestingly, a new
coupling $\gamma^*\gamma \,G^{(\vec k)}$ appears, allowing the
excited photons $\gamma^*$ to decay into gravitons and an ordinary
photon. This non-diagonal coupling is modulated by the 
``transition form factor'' $g_1^{(k_1,k_2)} + g_1^{(k_2,k_1)}$.
The coupling of $\gamma^*$ to fermions, which is so interesting in
view of the possibility of producing such a resonance in  $e^+e^-$
or $\mu^+\mu^-$ collisions, arises only if the fermions are located differently 
from the photons in the extra dimensions (this is the usual
assumption~\cite{Anton}, adopted in our fat-brane scenario).
In an alternative scenario, in which the fermions are also
spread over the width $L$ of the brane, as the bosons described
in Eq.~(\ref{photons}) are, the $e^+e^- \gamma^*$
coupling is forbidden by a KK momentum-conservation selection rule. 
For such boson--fermion symmetric {\it superfat branes,} 
the $\bar \psi \gamma^\mu \psi A_\mu^+$ coupling in
Eq.~(\ref{SQED}) should be removed\footnote{The Dirac algebra would be  
unchanged (no $\gamma_M$s are introduced), as in~\cite{sundrum1}.}.

Armed with the action of Eqs.~(\ref{SQED}) and (\ref{Tmunu}), one can proceed
to compute a phenomenologically interesting process, such as the production,
in an $e^+e^-$ or $\mu^+\mu^-$ collider, of single photons plus (invisible) 
gravitons: $\ell^+ \ell^- \to \gamma+E_{\rm mis}$. For the fat brane 
this process will be dominated by the $\gamma^*$ s-channel resonance. 
To facilitate the comparison with previous work on  thin branes by 
Giudice {\it et al.}~\cite{Giudice},
which we have checked to be correct, we shall concentrate on a hypothetical
$\sqrt{s}=1$ TeV collider, and present cross sections with
the same cuts as they use (a transverse energy cut 
$E_{T,\gamma}\equiv p_\gamma\,\sin\theta$ $\geq E_{T,\gamma}^{\rm min}$
to avoid collinear divergences and a cut $E_\gamma\leq$ 450 GeV to avoid
the dominant $\gamma\, Z\to\gamma \nu\bar \nu$ background).
In  a letter we cannot present a plethora of results
and we mainly discuss the  $\delta\!=\!2$ case, even
if it is presumably marginalized, for parameters in the range
we shall study, by astrophysical considerations~\cite{Supernova}.
Our main message --on how rich the phenomenology 
can be-- is  sufficiently illustrated by this case, 
though for the most interesting observables
we also work out results for $\delta=4,\,6$.
The parameters to be varied are the ``new-physics'' scale $M$ and the 
width of the brane $L$ or, equivalently, the mass of the $\gamma^*$, $m=2\pi/L$.
We do not know the exact relation between $m$ and $M$, which should be
of the same order of magnitude ($R$ is related to $M$ by 
$M_P^2\!=\![ 2 \pi R]^\delta M^{\delta + 2}$).

\begin{figure}[t]
\begin{tabular}{cc}
\hskip -0.75truecm 
\epsfig{file=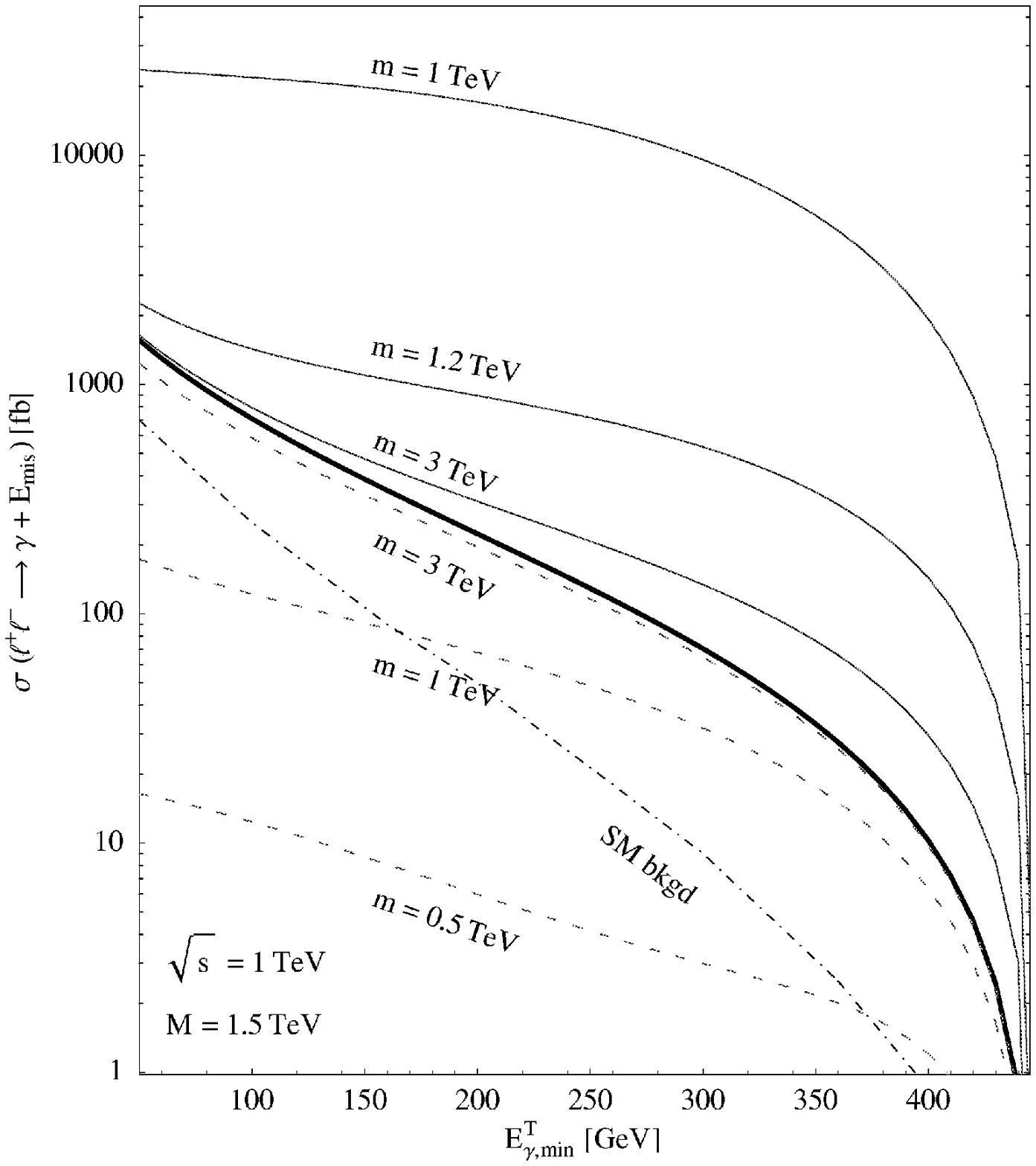, width=7.6cm} &
\hskip 0.2truecm 
\epsfig{file=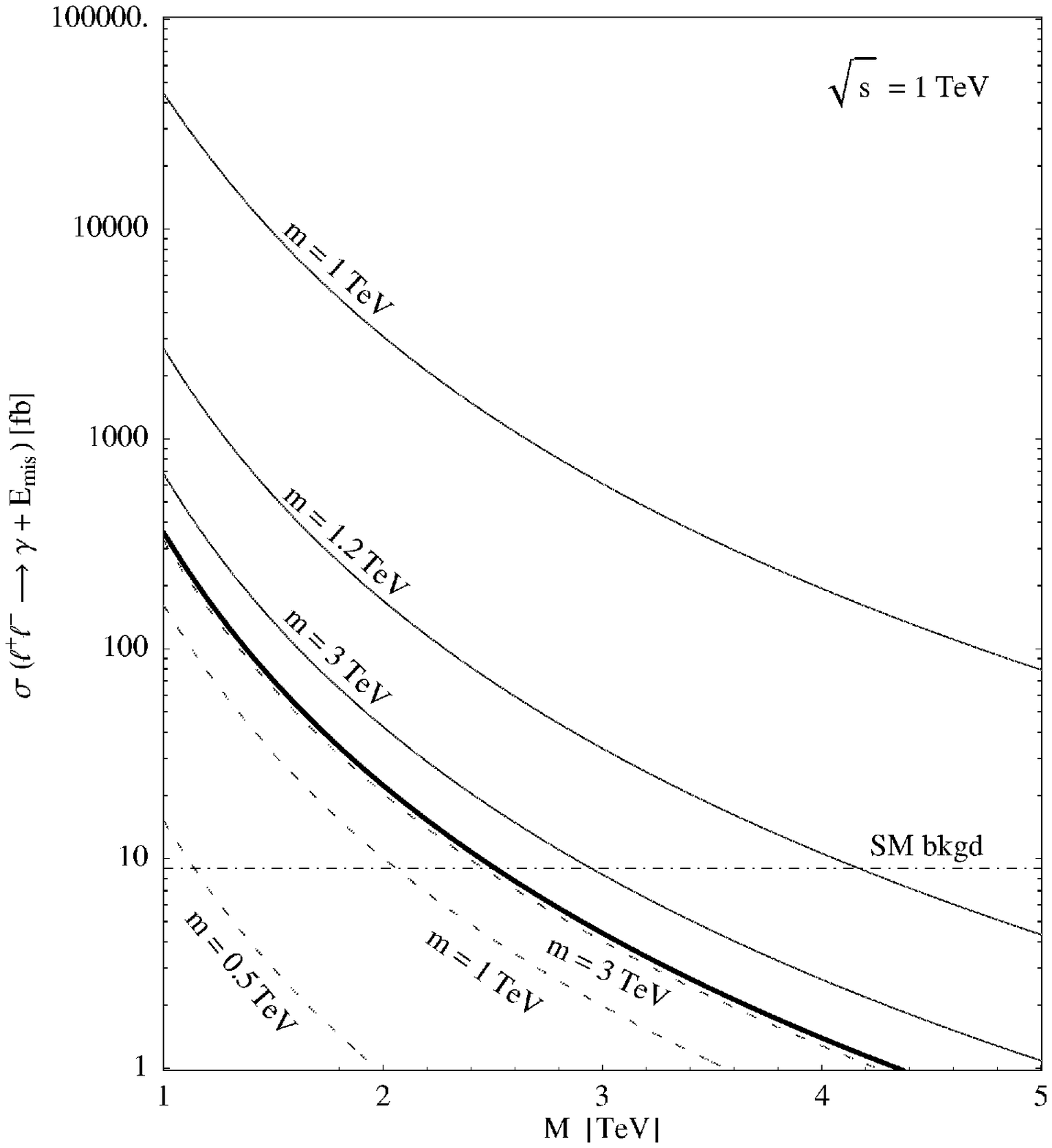, width=8cm} \\
\hskip -0.3truecm
{\small (a)}            &
\hskip 0.4truecm
{\small (b)}
\end{tabular}
\caption{{The cross sections for $\ell^+\ell^-\to \gamma+E_{\rm mis}$ at
$\sqrt{s}=1$ TeV, for $\delta\!=\!2$. (a) As a function of  
$ E_{T,\gamma}^{\rm min}$.
(b) Integrated for $E_{T,\gamma}\!>\! 300$ GeV. 
The thick line in a) and b) is for thin branes.
Curves above (below) it are for fat (superfat) branes.}}
\label{fig:giudice}
\end{figure}

From the point of view of experimental observability, fat and superfat 
branes turn out to be, respectively, the optimistic and pessimistic
extremes, while thin branes 
are in between. This can be seen in
Fig.~\ref{fig:giudice}, where we present results for a lepton collider 
running at $\sqrt{s}=1$ TeV. In
Fig.~\ref{fig:giudice}(a) we show the 
cross section for $\ell^+\ell^-\to \gamma+E_{\rm mis}$, integrated for
$E_{T,\gamma}\geq E_{T,\gamma}^{\rm min}$ for various values of
$m$ and a fixed $M=1.5$ TeV. The results scale as $M^{-4}$, as
can be explicitly seen in Fig.~\ref{fig:giudice}(b)  for the 
total cross section (within cuts) as a function of $M$.
In both of these figures the thick continuous line is the
thin-brane limit~\cite{Giudice}. The continuous curves  are 
fat-brane results, they peak for $m=1$ TeV, when our assumed collider 
is running at the $\gamma^*$ resonant peak. 
The dashed curves are for superfat branes (in the labels
of these curves $m$ is a shorthand for $2\pi/L$, the $\gamma^*$
exists, but 
is not an $\ell^+\ell^-$ resonance in this case). The dash-dotted line 
is the Standard Model background, as estimated in~\cite{Giudice}.
The signal over background ratios are very favourable, except
for superfat branes at small $m$, for which the form-factor effects
quench the signal very significantly.

\begin{figure}[t]
\begin{tabular}{cc}
\hskip -0.75truecm
\epsfig{file=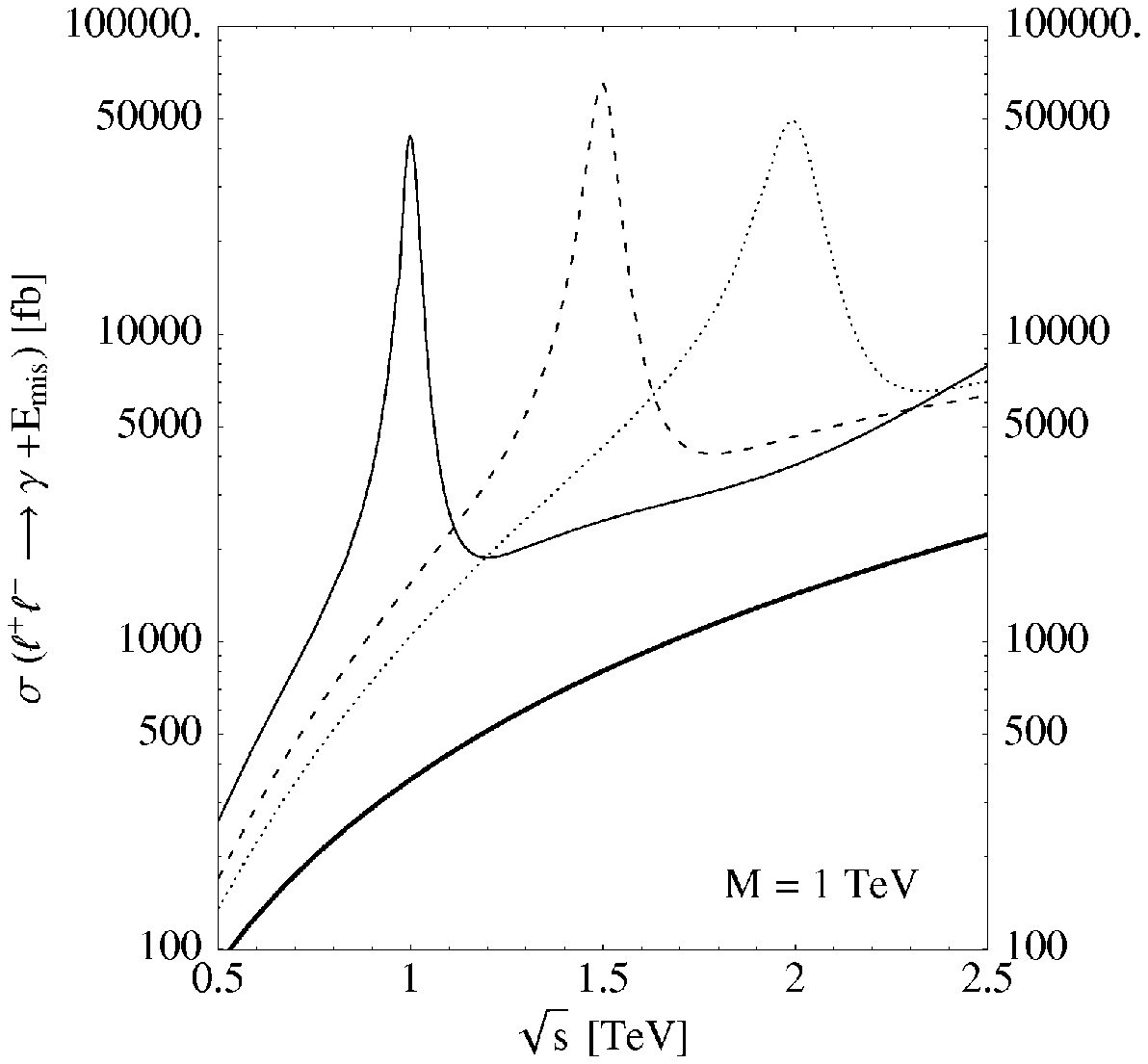, width=8.6cm} &
\hskip -0.3truecm
\epsfig{file=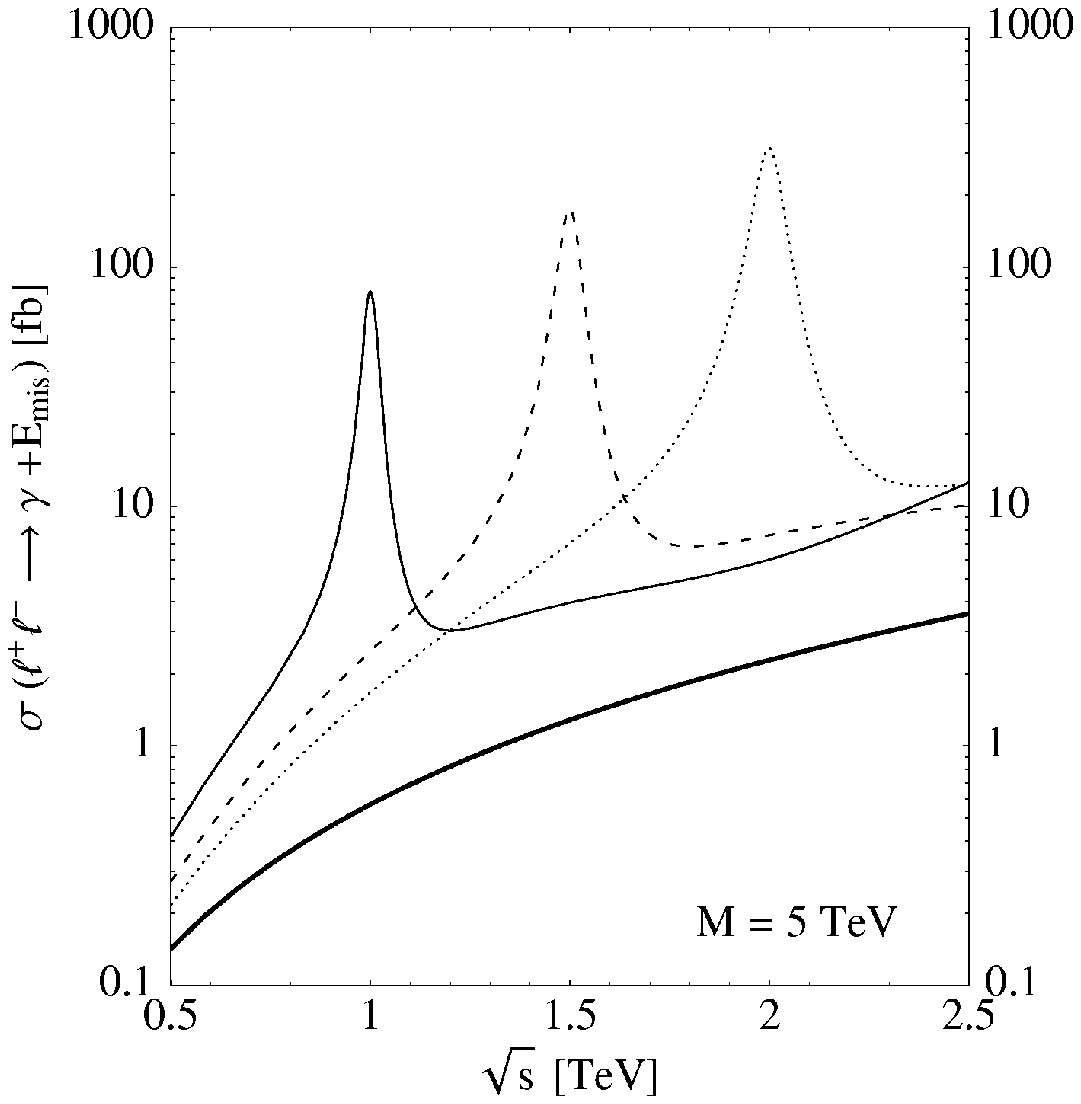, width=8cm} \\
\hskip 0.2truecm
{\small (a)}            &
\hskip 0.2truecm
{\small (b)}
\end{tabular}
\caption{{The fat-brane resonant cross sections for $\ell^+ \ell^- \to 
\gamma+E_{\rm mis}$ as a function of $\sqrt{s}$, for $M= 1$ TeV (a)
and $M= 5$ TeV (b), for $\delta=2$. The non-resonant curves are thin-brane
results.}}
\label{fig:resonances}
\end{figure}

In any model with branons,
KK recurrences, or any other sequential gauge bosons that couple to 
fermions, the $\ell^+\ell^-$ total annihilation cross section 
 would have the obvious resonant-peak signatures, 
no doubt interestingly intertwined~\cite{Anton,chorvos}, since
a $Z^*$ peak at $s=m^2+M_Z^2$ is to be expected, with $m$ the mass
of $\gamma^*$. The total cross section, however, would not distinguish the
objects we are discussing --related to bulk dimensions in which 
only gravitons dwell-- from  more conventional KK excitations.
We thus concentrate on a specific final state: $\gamma$ 
plus unobserved gravitons. In Fig.~\ref{fig:resonances} we show
$\sigma(\ell^+\ell^-\to\gamma+E_{\rm mis})$ as a function of $\sqrt{s}$ 
for the case of fat branes, for $m=1,\,1.5,\,2$ TeV and $M=1,\,5$ TeV, all for
$\delta=2$. Also shown, for comparison, is the thin-brane 
result~\cite{Giudice}.

The fat-brane results of Figs.~\ref{fig:giudice} and \ref{fig:resonances} 
depend on the total and partial widths of the
$\gamma^*$ resonance, which we now discuss in some detail. At tree level,
$\gamma^*$ decays into charged lepton pairs , hadrons ($q\bar q$ pairs),
photon plus gravitons ($\gamma \, G^{(\vec k)}$) and scalar photon plus gravitons
($\tilde \gamma\,G^{(\vec k)}$), an ``invisible'' decay channel. The widths into all
these channels can be worked out from the action of 
Eqs.~(\ref{SQED}) and (\ref{Tmunu}), and its generalization to fractionally
charged quarks. For the fermion channels we obtain
$\Gamma_{\gamma^*} (q\bar q) \simeq 1.6\times 10^{-2} \; m$ and
$\Gamma_{\gamma^*} (\ell^+\ell^-)\simeq 2.6\times 10^{-2} \; m $,
independent of $M$ and $\delta$. 

For the $\gamma^*\!\to\!\gamma\,G^{(\vec k)}$ decays, 
a fixed photon energy $E_\gamma$ 
corresponds to a given graviton mass $|\vec k|/R$ in their
quasi-continuum mass distribution. Define $x\!\equiv\! 2\,E_\gamma/m$,
so that $x=1-|\vec k|^2/(m R)^2$ and $0\!\leq\! x\!\leq\! 1$. 
The differential width is of the form:
\be
{d\,\Gamma_\delta\over dx}=
m\;\left({m\over M}\right)^{\delta+2}\;
{d\,\chi_\delta\over dx}\; ,
\label{widthscaling}
\ee
where the scaled width $d\chi_\delta/dx$ is independent of
$M$ and $m$.
For the case $\delta\!=\!2$, define  $\alpha\!\equiv\!\arctan (k_1/k_2)$
and let $g(x,\alpha)$ be a shorthand for $g_1^{(k_1,k_2)} + g_1^{(k_2,k_1)}$.
For the scaled width in Eq.~(\ref{widthscaling}), we get:
\be
{d\chi_2\over dx}={ M_P^2 \, x \over 96\,\pi}\;
\int_0^{2\pi}\,d\alpha\; \left \{ {|g|^2 \over 12}\;{x^2 \over (1-x)^2}\;
\left [ 10 - 15 x + 6 x^2 \right ] \right \}\; ,
\label{diffwidth}
\ee
where the quantity in wiggly brackets is the square of the
transition matrix element for $\gamma^*\!\to\!\gamma\,G^{(\vec k)}$
decay and depends, via $g$, on the form factors specific to
a particular model of confinement to a brane. 
In our model, for the total ($x$-integrated) width 
$\gamma^* \rightarrow \gamma\,G $, we obtain 
$\chi_2 = 1.1 \times 10^{-3}$, to be substituted in Eq.~(\ref{widthscaling}).
The $\gamma^* \rightarrow \tilde\gamma\,G $ width has the same dependence
on $m$ and $M$, with a different coefficient, $\tilde \chi_\delta$.
For $\delta = 2$, $\tilde \chi_2 = 5.7 \times 10^{-4}$.
The generalization of
Eq.~(\ref{diffwidth}) to $\delta\! > \! 2$ is rather straightforward~\cite{us}. 
The integrated widths for $\gamma^*\!\to\!\gamma\,G$
are given by Eq.~(\ref{widthscaling}), with $\chi_4 = 1.0$ and $\chi_6 = 7.4 \times 10^2$.
The rapid growth of $\chi_\delta$ with $\delta$ reflects 
how fast the number of accessible KK excitations increases, as new dimensions 
are added.

\begin{figure}[t]
\begin{tabular}{cc}
\hskip -0.75truecm
\epsfig{file=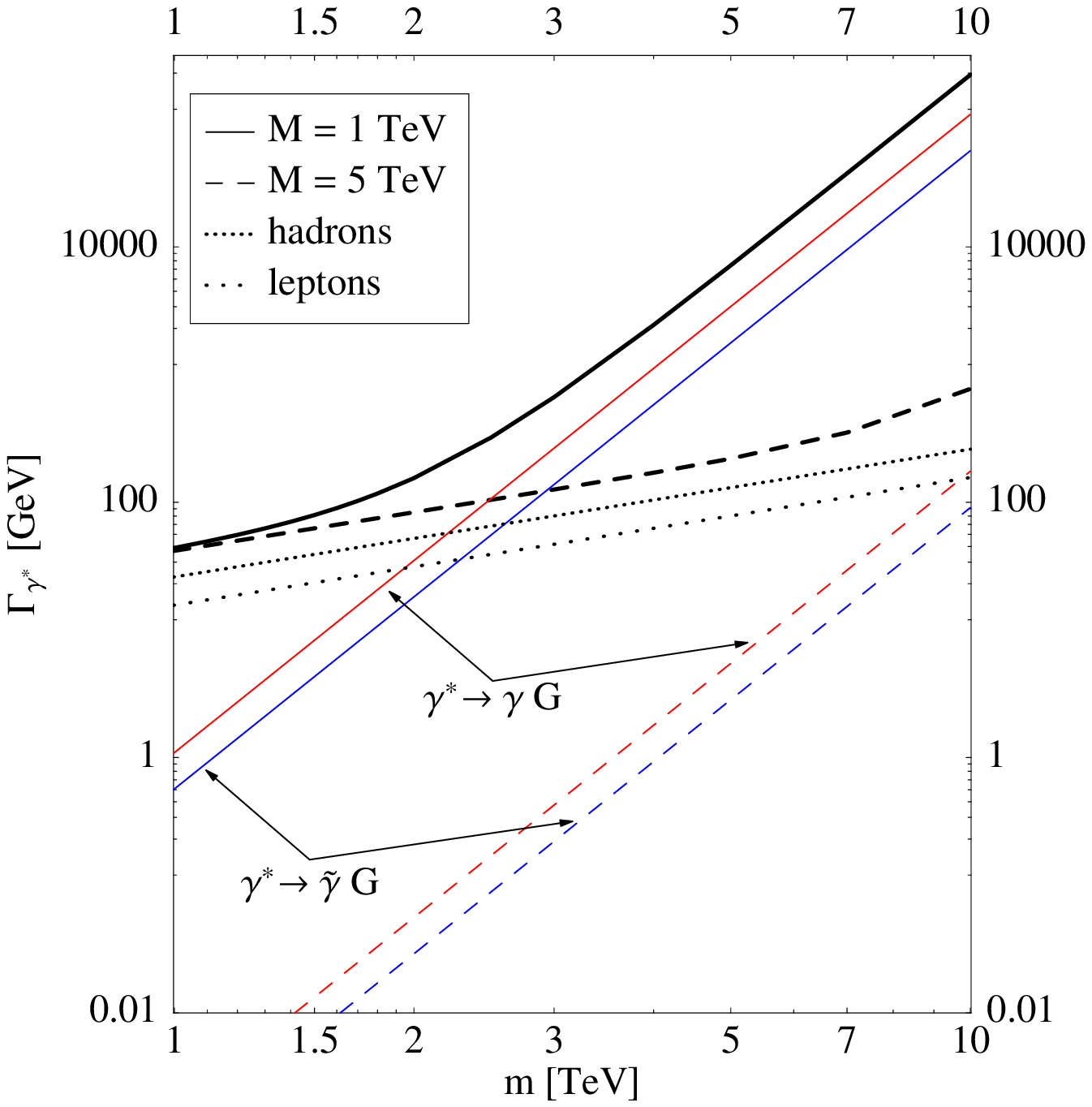, width=8.55cm} &
\hskip -0.5truecm
\epsfig{file=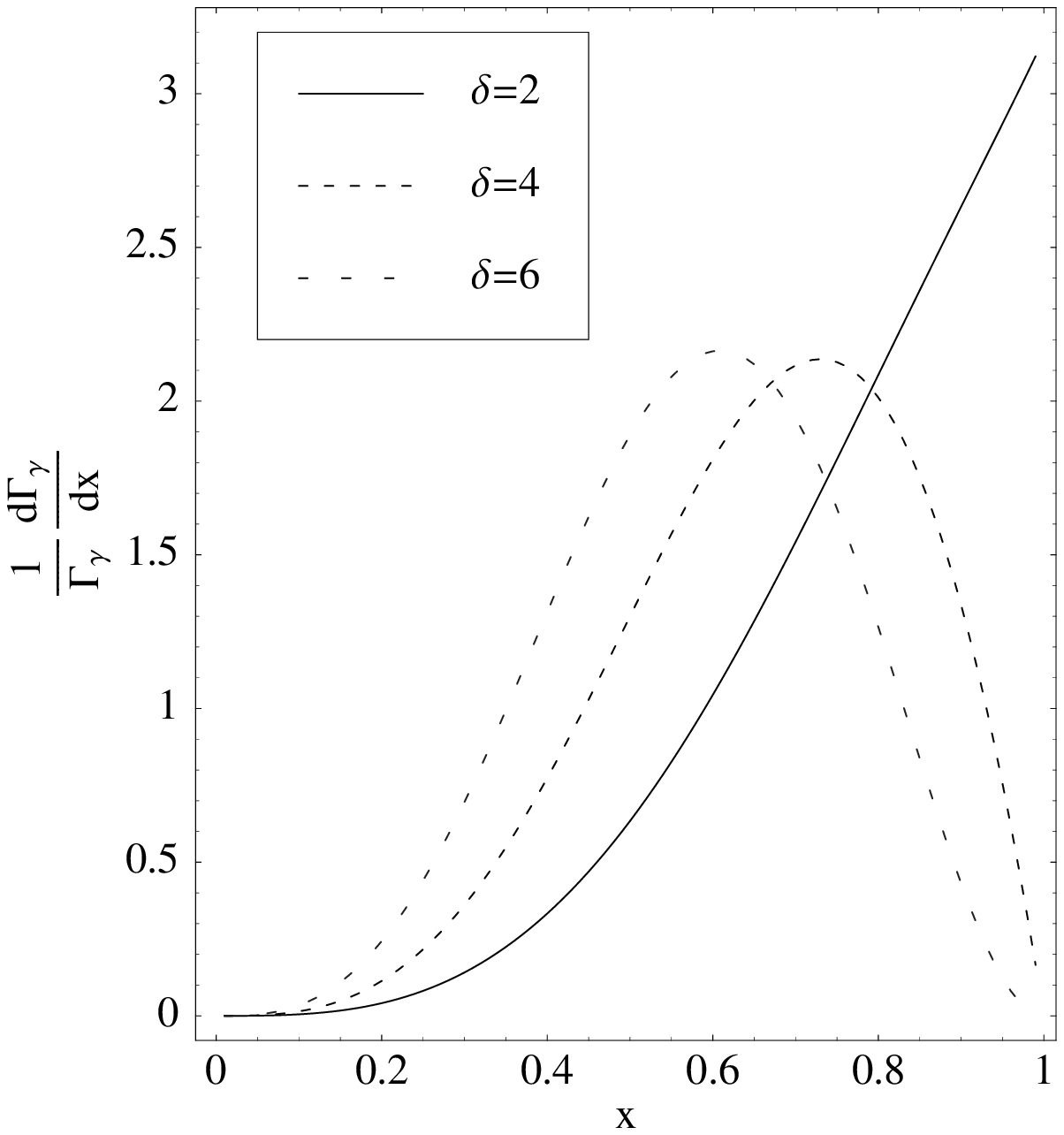, width=7.75cm} \\
\hskip -0.2truecm
{\small (a)}            &
\hskip 0.3truecm
{\small (b)}
\end{tabular}
\caption{{ (a) Partial and total widths of the $\gamma^*$ resonance, as 
functions of its mass, $m$, for $\delta=2$. The continuous (dashed) curves 
are for $M=1,\,5$ TeV. The dotted lines are partial widths into quarks and 
leptons. (b) The decay spectrum $d\Gamma_\gamma/(\Gamma_\gamma\,dx)$ 
for $\gamma^*\to\gamma + E_{\rm mis}$ as a function of $x$, for three 
values of $\delta$.}}
\label{fig:widths}
\end{figure}

In Fig.~\ref{fig:widths}(a) we show the total and partial widths of the 
$\gamma^*$ as functions of $m$, for $M\!=\!1,\,5$ TeV and $\delta\!=\!2$.
For the smaller and most accessible $\gamma^*$ masses the partial width 
into fermions dominates. But even at these relatively small $m$, for $M$ 
also of ${\cal O}(1)$ TeV, the $\gamma^*\to \gamma\,G$ branching ratio 
is large enough to constitute a striking signal. 
Only for $M$ large relative to $m$ do the branching ratios into this
tell-tale channel become unfavourably small. For $ \delta > 2 $ the
resonances get wider: they remain prominent peaks
only for sufficiently small $m/M$, 
as dictated by Eq.~(\ref{widthscaling}).

In Fig.~\ref{fig:widths}(b) we show the most spectacular fat-brane signal: 
the spectrum of photon energies in $\gamma^*\!\to\!\gamma\, G$ decay. 
This distribution is
a superposition of the peaks associated with the individual gravitons
of different mass, but their number is so huge as to make
their resolution impossible. The scaled function
$d\Gamma_\gamma/(\Gamma_\gamma\,dx)$ 
is independent of $M$ and $m$; it only depends on 
$\delta$ and on the explicit branon wave functions, 
of which Eq.~(\ref{photons}) is an example. 
Only the $\delta$-dependence survives
as $x$ approaches unity, since this region corresponds
to decays into the lighter gravitons, for which the power behaviour of the
form factors of Eqs.~(\ref{diagonal}) and (\ref{nondiagonal}) and Fig.~\ref{fig:pllog0ik} 
is model-independent: the matrix element in
Eq.~(\ref{diffwidth}) and its generalizations to $\delta\!>\!2$ is a
polynomial~\cite{us} in $x$  ($|g|\!\propto\! ( 1\! -\! x)^2$ as $x\!\to\! 1$).
Close to $x=1$ we obtain:
\be
\frac{1}{\Gamma_\gamma}\frac{d \Gamma_\gamma}{d x} \propto
(1 - x )^{(\delta-2)/2}\;\left[1+{\cal O}\left([1-x]\right)\right]
\ee

A two-body
decay into a $\gamma$ and a single invisible particle would result in a
peak with a width governed by resolution, looking nothing like the curves
of Fig.~\ref{fig:widths}(b). A three-body decay involving two invisible
particles --with form factors contrived to imitate one of the shapes
in the figure-- is the only implausible impostor. 
Thus, the measurement of $d\Gamma_\gamma/(\Gamma_\gamma\,dx_\gamma)$ 
would be a convincing signal for the existence of extra dimensions,
and its behaviour as $x$ approaches its upper limit would 
constitute a very neat {\it dimensiometer}.

To summarize, the message conveyed by Figs.~\ref{fig:giudice}--\ref{fig:widths} 
is clear: the physics at energies
at which new dimensions ``open up'' can range from the very
challenging to the very rich. The phenomenological
signals so far studied in the literature~\cite{Giudice,Peskin1}
 can be wiped-out by ``form-factor''
effects if branes are ``superfat'' and have widths of the natural order of magnitude.
On the other hand, if chiral fermions are special and localized
in slices of space-time thinner than those in which bosons reside 
--as for the more standard fat branes-- the 
production of KK-like ``branon'' resonances in collisions between 
ordinary particles becomes possible. The decays of 
excited $\gamma$'s or $Z$'s into their ground states plus unobservable
gravitons would provide astounding signals. In particular, the energy
distribution of the final-state $\gamma$ or $Z$ would not be sharply
peaked (as for a conventional two-body decay), its spread reflecting
the mass distribution of the accompanying tower of gravitons,
and its high-energy tail providing a direct measurement of $\delta$, 
the {\it number} of large extra dimensions.

At high energy, ideas about large extra dimensions will first confront
new data at the LHC $pp$ collider, as we plan to discuss in
subsequent work. In the fat-brane scenario, the
$q\bar q$ production
of $Z^*$ --the first branon excitation of the $Z$-- and its decay into $Z$ plus KK
gravitons, constitute the most spectacular signatures. 
The energy distribution of the observed $Z$'s, because of the unobservable
longitudinal momentum of the colliding partons, is not as gorgeous a signal
as its lepton-collider counterpart. But this inconvenience is compensated
by large statistics and, what's more, by the LHC's appropinquity.

\section*{Acknowledgements}
We kindly thank E.~Alvarez, K.~Benakli, A.~Cohen, S.~Dimopoulos, G.~Dvali, 
F.~Feruglio, C.~G\'omez, L.~E.~Ib\'a\~nez, E.~Lopez, W.~Lerche, 
J.~March-Russell, C.~Mu\~noz, T.~Ortin, R.~Sundrum and F.~Zwirner
for comments and discussions. 
A.~D. acknowledges the I.N.F.N. for financial support. 
S.~R. acknowledges the European Union for financial support 
through contract ERBFMBICT972474 and the Dept. of Physics of the University
of Michigan, Ann Arbor. The work of A.~D., M.~B.~G. 
and S.~R. was partially supported by the CICYT project AEN/97/1678.

{\bf Note added.}
Dudas and Mourad and Cullen {\it et al.}~have recently 
posted~\cite{Dudas:1999gz,Peskin2}, in which 
extra-dimensional signatures are studied in a complementary 
framework --stringy dynamics-- that also goes beyond the low-energy 
effective considerations of~\cite{Giudice,Peskin1}.

%
%
%

%

%

\begin{thebibliography}{99}
%
\bibitem{KK} 
T.~Kaluza, 
Sitzungsberichte of the Prussian Acad. of Sci. (1921) 966;\\
O.~Klein, 
Z. Phys. {\bf 37} (1926) 895.
%
\bibitem{RSh} 
V.~Rubakov and M.~Shaposhnikov, 
Phys. Lett. {\bf B 125} (1983) 136.
%
\bibitem{HW} 
P.~Horava and E.~Witten, 
Nucl. Phys. {\bf B 460} (1996) 506; 
Nucl. Phys. {\bf B475} (1996) 94.
%
\bibitem{Antoniadis:1990ew}
I.~Antoniadis,
Phys.\ Lett.\  {\bf B246} (1990) 377.
%
\bibitem{Savas} 
N.~Arkani-Hamed, S.~Dimopoulos and G.~Dvali, 
Phys. Lett. {\bf B 429} (1998) 263;
I.~Antoniadis, N.~Arkani-Hamed, S.~Dimopoulos and G.~Dvali, 
Phys. Lett. {\bf B 436} (1998) 257;
%
\bibitem{Lykken} 
J.~D.~Lykken, Phys. Rev. {\bf D54} (1996) 3693.
%
\bibitem{Review} 
For a recent review see for example T.~Banks, M.~Dine and A.~Nelson, 
JHEP {\bf 9906} (1999) 014 and reference therein.
%
\bibitem{Polchinski}
J.~Polchinski, hep-th/9611050.
%
\bibitem{Ibanez} 
G.~Aldazabal, A.~Font, L.~E.~Ib\'a\~nez and G.~Violero, 
Nucl. Phys. {\bf B536} (1998) 29; 
L.~E.~Ib\'a\~nez, C.~Mu\~noz and S.~Rigolin, 
Nucl. Phys. {\bf B553} (1999) 43;
G.~Aldazabal, L.~E.~Ib\'a\~nez and F.~Quevedo, 
JHEP {\bf 0001} (2000) 31 and hep-ph/0001083.
%
\bibitem{Dienes}
K.~R.~Dienes, E.~Dudas and T.~Gherghetta, 
Nucl. Phys. {\bf B537} (1999) 47.
%
\bibitem{Arkani} 
N.~Arkani-Hamed and M.~Schmaltz, 
Phys. Rev. {\bf D61} (2000) 033005;
N.~Arkani-Hamed, Y.~Grossman and M.~Schmaltz, hep-ph/9909411.
%
\bibitem{Bando} 
This point has also been investigated, though very differently, in
M.~Bando, T.~Kugo, T.~Noguchi and K.~Yoshioka, 
Phys. Rev. Lett. {\bf 83} (1999) 3601; 
J. Hisano and N. Okada, hep-ph/9909555;
T.~Kugo and K.~Yoshioka, hep-ph/9912496.
%
\bibitem{sundrum1}
R.~Sundrum, 
Phys. Rev. {\bf D 59} (1999) 085009.
%
\bibitem{us} 
A detailed discussion will appear in a paper in preparation
by the present authors. 
%
\bibitem{add2}
N.~Arkani-Hamed, S.~Dimopoulos and G.~Dvali, 
Phys. Rev. {\bf D 59} (1999) 086004.
%
\bibitem{Giudice} 
G.~Giudice, R.~Rattazzi and J.~Wells, 
Nucl. Phys. {\bf B 544} (1999) 3.
%
\bibitem{Peskin1}
E.~A.~Mirabelli, M.~Perelstein and M.~E.~Peskin, 
Phys. Rev. Lett. {\bf 82} (1999) 2236; \\
T.~Han, J.~D.~Lykken and R.-J.~Zhang, 
Phys. Rev. {\bf D 59} (1999) 105006.
%
\bibitem{Anton} 
I.~Antoniadis, C.~Mu\~noz and M.~Quir\'os, Nucl. Phys. {\bf B397} (1993) 515;
I.~Antoniadis, K.~Benakli and M.~Quir\'os, Phys. Lett. {\bf B331} (1994) 313,
and references therein.
%
\bibitem{Supernova} 
S.~Cullen and M.~Perelstein, 
Phys. Rev. Lett. {\bf 83} (1999) 268.
For cosmological limits, see 
L.~J.~Hall and D.~Smith, 
Phys. Rev. {\bf D60} (1999) 085008.
%
\bibitem{chorvos}
A.~Donini and S.~Rigolin, 
Nucl. Phys. {\bf B550} (1999) 59 and hep-ph/9905293;
I.~Antoniadis, K.~Benakli and M.~Quir\'os, 
Phys. Lett.  {\bf B460} (1999) 176;
E.~Accomando, I.~Antoniadis and K.~Benakli, hep-ph/9912287.
%
\bibitem{Dudas:1999gz}
E.~Dudas and J.~Mourad,
hep-th/9911019.
%
\bibitem{Peskin2}
S.~Cullen, M.~Perelstein and M.~E.~Peskin, hep-ph/0001166.
%
\end{thebibliography}
\end{document}